\documentclass[aps,physrev,preprint,amsmath]{revtex4-2}
\usepackage{graphicx}
\newcommand{\vv}[1]{\mathbf{#1}}
\renewcommand{\d}[1]{\ensuremath{\operatorname{d}\!{#1}}}

\begin{document}

\title{Coarse-graining to create minimalist models for dynamic, end-linked star-polymer networks}

\author{Tyla R. Holoman}
\affiliation{McKetta Department of Chemical Engineering, The University of Texas at Austin, Austin, Texas}

\author{C. Levi Petix}
\affiliation{Department of Chemical Engineering, Auburn University, Auburn, Alabama 36849, USA}

\author{Michael P. Howard}
\email{mphoward@auburn.edu}
\affiliation{Department of Chemical Engineering, Auburn University, Auburn, Alabama 36849, USA}

\author{Thomas M. Truskett}
\email{truskett@umich.edu}
\affiliation{Department of Chemical Engineering and Biointerfaces Institute, University of Michigan, Ann Arbor, Michigan}
\affiliation{McKetta Department of Engineering, The University of Texas at Austin, Austin, Texas}

\date{\today}

\begin{abstract}
Although minimalist models for patchy attractive particles have revealed powerful design rules for how particle valence directs colloidal assembly, less work has focused on comparably simple models of reversible, network-forming star polymers. Here, we use relative-entropy coarse-graining and simulation results from finer-resolution bead--spring star polymers to generate 5-bead models of dynamic, end-associating four-armed poly(ethylene glycol) macromers. Our results comparing structural correlations, network connectivity, and phase behavior of the coarse- and fine-grained models provide insight into how the accuracy and transferability of the coarse-grained models depend on the state point chosen for coarse-graining. The results also reveal intrinsic trade-offs between reducing degrees of freedom and expanding the range of effective interactions in coarse-graining that impact the total number of pairwise interactions in the resulting model, with implications for its computational efficiency. 
\end{abstract}

\maketitle

\section{Introduction \label{intro}}
Gels are solvent-filled, percolated networks of colloidal or polymeric building blocks. Materials in gel states can exhibit multiple application-enabling properties, including mechanical responsiveness, processability, and self-healing, though attaining high performance in one or more of these areas often compromises performance in others. As a result, considerable attention has focused on understanding the performance trade-offs and their microscopic origins~\cite{zaccarelli2007colloidal,Shibayama2018Gels:BioMatter,petekidis2021,green2022,tang2021dynamic,webber2022dynamic,Bertsch2023Self-HealingRegeneration,lee2025dynamic}.

Conventional polymer gels are prepared by randomly connecting linear polymers into a three-dimensional network with small-molecule crosslinkers, creating an inhomogeneous gel structure~\cite{shibayama1998spatial}. In contrast, this work focuses on a class of networks formed by a different strategy: reversibly linking star-polymer building blocks (i.e., macromers) with a small number of arms that are short enough so that entanglement effects on gel properties can be safely ignored. Bond reversibility and specificity in these materials can be ensured by forming the gel network from a binary mixture of stars with arms that are end-functionalized with complementary dynamic covalent bonding pairs~\cite{Sakai2008DesignMacromonomers,Yesilyurt2016InjectableProperties,Yesilyurt2017MixedNetworks,Apostolides2017DynamicMaterials,Parada2018IdealNetworks,Shibayama2019PrecisionFuture,Ahmadi2020DynamicCrosslinks,Fitzsimons2020PreferentialAdditions,FitzSimons2022EffectBonds,crowell2023shear,crowell2026leveraging}. Gel networks assembled this way have homogeneous structures~\cite{Sakai2008DesignMacromonomers,matsunaga2009sans,matsunaga2009structure,akagi2011examination} and complex rheological behavior that can be tuned by selecting solution conditions (e.g., pH) or the chemistry of the dynamic bonds~\cite{wang2015adaptable,ollier2023biomimetic,zheng2024real,sing2015celebrating,mahmad2020understanding,Ahmadi2020ThermodynamicCoordination,Ahmadi2021CoordinationNetworks,Fitzsimons2020PreferentialAdditions,FitzSimons2022EffectBonds,crowell2023shear,crowell2026leveraging}.

To understand and design these gels, we need models that relate their macroscopic behavior to the molecular properties of the stars and the reactivity of their dynamic bonds. Given the intrinsic multiscale nature of gel networks, coarse-resolution models serves as an important component in understanding how changes in star architecture, or in the thermodynamics and kinetics of arm-linking, affect emergent properties such as network structure, phase behavior, and rheology~\cite{Parada2018IdealNetworks,raffaelli2021stress,kumar2023reversible,Holoman2025SimulatingMaterials,vigil2025coherent,miotti2026mesoscopic}. Insights from coarse-grained models developed for these materials may also reveal relationships across gel network classes. For example, coarse-grained models have revealed design rules for colloidal particles that form limited-valence networks by aligning attractive patches on their surfaces~\cite{zhang2005self,Bianchi2006PhaseLiquids,sciortino2011reversible,van2012formation,smallenburg2013liquids,reinhardt2013computing,Bianchi2015Soft-patchySelf-organization,liu2016diamond,lindquist2016formation,russo2022physics}. Beyond colloids, these patchy-particle design rules can predict the experimental phase behavior of associating DNA nanostars whose arms are rigid because of the long persistence length of a DNA double helix~\cite{biffi2015equilibrium,Rovigatti2014AccurateNanostars,rovigatti2014gels,Lattuada2021SpatiallyGels,conrad2022emulsion}. 

In this work, we use systematic coarse-graining to extend our understanding of reversible gel formers to a wider class of network-forming star polymers with flexible arms, focusing on tetra-functional poly(ethylene glycol), or tetra-PEG. To create a simple representation of tetra-PEG that captures the effects of arm flexibility, we introduce a 5-bead (4-arm plus one central junction) model and determine its model parameters by minimizing the relative entropy~\cite{shell:jcp:2008, Shell2016COARSEGRAININGENTROPY, Sreenivasan2024Relentless:Optimization} against a finer-grained bead--spring model for star polymers~\cite{Kremer1990DynamicsSimulation, Furuya2020MolecularProperties} in the absence of dynamic inter-star bonding. To validate the 5-bead model, we compare its predictions of network bonding motifs and propensity for liquid--liquid phase separation with those of the finer-grained bead--spring model, where end-association between star arms in both models is treated using a recently introduced dynamic bonding scheme by Hocky and co-workers~\cite{Mitra2022ABinders}. Based on these results, we identify key aspects of tetra-PEG network behavior that the 5-bead model accurately reproduces, as well as coarse-graining challenges for star polymer networks related to parameter transferability and computational efficiency. This information should help guide which questions simple, coarse-grained representations can answer, while also indicating where more sophisticated coarse-graining strategies or more detailed modeling are needed to produce reliable predictions.

\section{Models and Methods}
\label{sec:methods}
\subsection{Models}
\begin{figure}
\includegraphics[width=\textwidth]{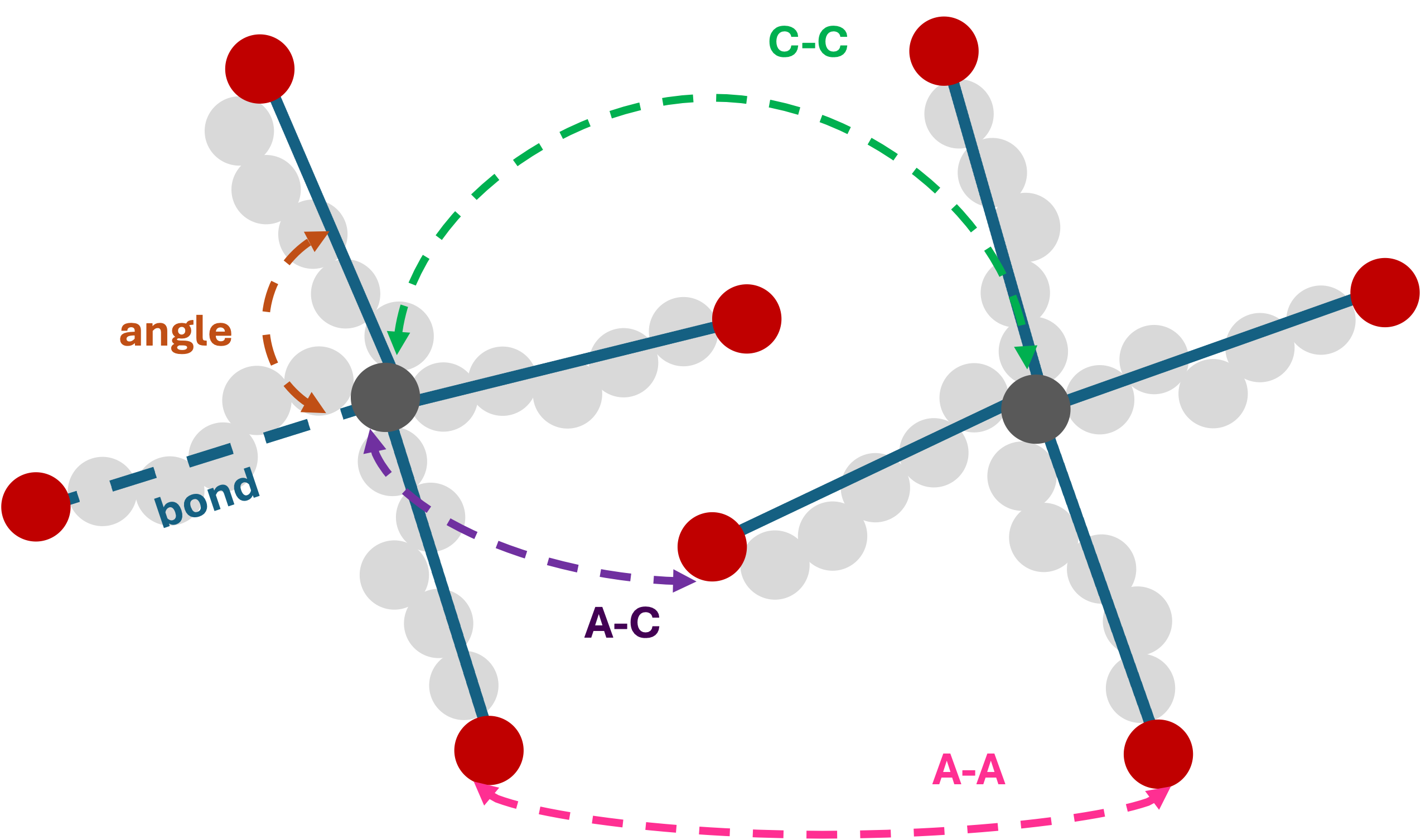}
\caption{{\bf{Schematic of fine-grained (FG) and coarse-grained (CG) models for tetra-PEG.}} Light gray circles represent the beads of the FG model, while dark gray circles and red circles represent the center (C) and arm (A) beads of the CG model, respectively. Designed interactions of the CG model are also depicted, which include the intramolecular angle and bond potentials and intermolecular center--center (C--C), arm--center (A--C), and arm--arm (A--A) potentials.}
\label{fig:model}
\end{figure}

We simulated network formation of tetra-PEG macromers using two star-polymer models with different resolutions, which we refer to as the fine-grained (FG) and coarse-grained (CG) models (Fig.~\ref{fig:model}). The FG model comprised four linear arms of six beads each that were bonded to a central bead \cite{Kremer1990DynamicsSimulation, Furuya2020MolecularProperties}, so each star polymer contained a total of 25 FG beads. The FG beads all had diameter $\sigma$ and mass $m$, and their interactions were modeled using the Kremer--Grest approach \cite{Kremer1990DynamicsSimulation}. Specifically, all FG beads interacted with each other pairwise through the purely repulsive Weeks--Chandler--Andersen (WCA) \cite{weeks:jcp:1971} potential
\begin{equation}
u_{\rm WCA}(r) = \begin{cases}
\displaystyle 4 k_{\rm B} T \left[\left(\frac{\sigma}{r}\right)^{12} - \left(\frac{\sigma}{r}\right)^6 + \frac{1}{4}\right], & r \le 2^{1/6}\sigma \\
0, & {\rm otherwise}
\end{cases},
\label{eq:wca}
\end{equation}
where $r$ is the distance between two beads, $T$ is the temperature that was held constant in all simulations, and $k_{\rm B}$ is the Boltzmann constant. Additionally, bonds between beads were represented by a finitely extensible nonlinear elastic (FENE) potential,
\begin{equation}
u_{\rm FENE}(r) = -\frac{1}{2} k r_0^2\ln\left[1-\left(\frac{r}{r_0}\right)^2\right],
\end{equation}
where $k = 30\,k_{\rm B}T/\sigma^2$ and $r_0 = 1.5\,\sigma$ are commonly used values for the Kremer--Grest spring constant and maximum bond length.

The CG model of tetra-PEG consisted of one central bead, denoted C, which was bonded to four arm beads, denoted A. The mapping $\mathcal{M}$ transformed each FG configuration into a CG configuration by assigning the FG center bead to the CG center bead and the last bead of each FG star arm to the CG arm bead. During model parameterization, the C and A beads had unit mass $m$ because only equilibrium structures were needed, but for production simulations with the CG model, the C and A beads had mass $5\,m$ so the total polymer mass was the same in both the FG and CG models. The intramolecular interactions consisted of a C--A bond potential and an A--C--A angle potential. The intermolecular interactions consisted of C--C, C--A, and A--A pair potentials between beads in different polymers. The parameters of the CG interaction potentials, denoted collectively by $\boldsymbol{\theta}$, were determined by bottom-up coarse-graining \cite{shell:jcp:2008, Shell2016COARSEGRAININGENTROPY} as described in Sec.~\ref{sec:methods:cg}. Readers less interested in these details can advance to Sec.~\ref{sec:methods:network} for a description of how network formation was simulated for both models.

\subsection{Coarse-grained model parametrization}
\label{sec:methods:cg}
We parametrized the CG model based on FG-model simulations where no star--star bonding was allowed to occur. Each FG simulation included 1000 stars in a cubic simulation box with periodic boundary conditions. Simulations were performed using Langevin dynamics in HOOMD-blue (version 3.8.0) with a timestep $\Delta t = 0.001 \tau$ and friction coefficient $0.1 \,m/\tau$, where $\tau = \sqrt{m\sigma^2/(k_{\text{B}}T)}$ was the unit of time. The stars were initialized in a $200\,\sigma^3$ cubic box and allowed to equilibrate for $2 \times 10^3\,\tau$ before the box was compressed to attain a target density using a volume ramp that scales the box lengths at a constant rate over $2 \times 10^3\,\tau$. The stars were then equilibrated for another $2 \times 10^3\,\tau$, followed by a production run where configurations were recorded every $50\,\tau$ for $2\times 10^4\,\tau$. 

\begin{figure}
 \includegraphics[]{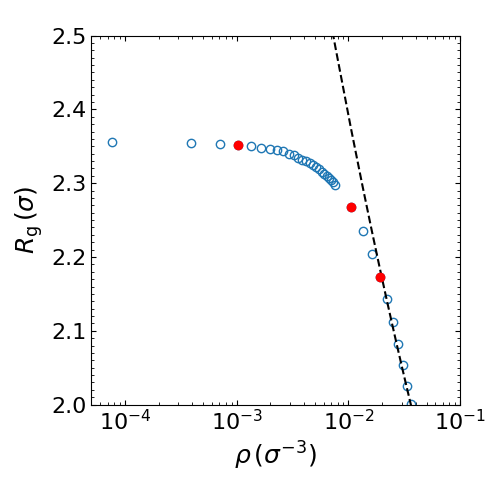}
 \caption{\label{fig:rgs} {\bf{Crossover from dilute to semidilute conditions}}. Radius of gyration for the FG tetra-PEG model versus star number density $\rho$. Red points represent the three selected conditions for coarse-grained model parametrization.}
 \end{figure}

We chose three concentrations that we refer to as dilute, intermediate, and semidilute for parameterizing the CG model based on FG simulations. These concentrations are indicated as red filled circles in Fig.~\ref{fig:rgs}, which shows how the root mean squared radius of gyration $\langle R_{\rm g}^2 \rangle^{1/2}$ of the FG model depends on the star-polymer number density $\rho$ (Fig.~\ref{fig:rgs}). In the semidilute regime, the radius of gyration is expected to scale with $\rho^{-0.125}$ \cite{Rubenstein2003PolymerPhysics,Asai2013CorrelationGels}, and in our FG simulations, the radius of gyration scales as $\rho^{-0.11}$, approximately matching the theoretical expectation. We chose the dilute concentration to be at a density $\rho = 0.002\,\sigma^{-3}$, before $\langle R_{\rm g}^2 \rangle^{1/2}$ significantly decreases below its value at infinite dilution. The semidilute condition was chosen to be at $\rho=0.019\,\sigma^{-3}$, the beginning of the scaling regime, and the intermediate condition was assigned between these two with a number density $\rho = 0.011\,\sigma^{-3}$. 

The parameters of the interaction potentials were determined by minimizing the relative entropy $S_{\rm rel}$ between an ensemble of structures generated by the FG model and an ensemble of structures generated by the CG model \cite{shell:jcp:2008, Shell2016COARSEGRAININGENTROPY}, 
\begin{equation}
S_{\mathrm{rel}}(\boldsymbol{\theta}) = \int p_{\rm FG}(\vv{r}^N)\ln \left[\frac{p_{\rm FG}(\vv{r}^N)}{p_{\rm CG}(\mathcal{M}(\vv{r}^N);\boldsymbol{\theta})}\right] \d{\vv{r}^N} + S_{\rm map},
\label{eq:S}
\end{equation}
where $p_{\rm FG}$ is the probability density of observing a configuration of $N$ particles $\vv{r}^N$ in the FG ensemble, $p_{\rm CG}$ is the probability density of observing the mapped configuration $\mathcal{M}(\vv{r}^N)$ in the CG ensemble, and $S_{\rm map}$ is the mapping entropy. Of these quantities, only $p_{\rm CG}$ depends on the parameters $\boldsymbol{\theta}$ of the CG interaction potentials. For the canonical (isothermal--isochoric) ensemble of interest to this work, eq.~\eqref{eq:S} becomes
\begin{equation}
    S_{\mathrm{rel}}(\boldsymbol{\theta}) = \beta \left\langle U_{\rm CG}(\boldsymbol{\theta}) - U_{\rm FG} \right\rangle_{\rm FG} + \ln Z_{\rm CG}(\boldsymbol{\theta}) - \ln Z_{\rm FG} + S_{\rm map},
    \label{eq:srel_gen}
\end{equation}
where $\beta = 1/(k_{\rm B} T)$, $U_{\rm CG}$ and $U_{\rm FG}$ are the total potential energy of the CG and FG models for a given FG configuration, $\langle \cdot \rangle_{\rm FG}$ denotes an average over the FG ensemble, and $Z_{\rm CG}$ and $Z_{\rm FG}$ are the canonical configurational partition functions for the CG and FG models. The quantities that depend on $\boldsymbol{\theta}$ are explicitly denoted. Direct evaluation of eq.~\eqref{eq:srel_gen} is not usually possible because of the partition functions; however, differentiating with respect to $\boldsymbol{\theta}$ gives
\begin{equation}
    \frac{\partial S_{\rm rel}}{\partial \boldsymbol{\theta}} = \beta \left\langle \frac{\partial U_{\rm CG}}{\partial \boldsymbol{\theta}}\right\rangle_{\rm FG} - \beta \left\langle \frac{\partial U_{\rm CG}}{\partial \boldsymbol{\theta}}\right\rangle_{\rm CG},
    \label{eq:gradgen}
\end{equation}
where $\langle \cdot \rangle_{\rm CG}$ denotes an average over the CG ensemble. The average over the FG ensemble can be evaluated for a given $\boldsymbol{\theta}$ by computing the derivatives of $U_{\rm CG}$ on configurations from a reference FG simulation mapped to CG configurations. The average over the CG ensemble can be evaluated for a given $\boldsymbol{\theta}$ by performing a simulation with the CG model to generate configurations. This gradient can be used to minimize $S_{\rm rel}$ with respect to $\boldsymbol{\theta}$ using iterative methods \cite{Sreenivasan2024Relentless:Optimization}.

The intramolecular CG interactions were represented using Akima splines having 24 knots uniformly spaced for bond lengths between $0.9\,\sigma$ and $6\,\sigma$ and 13 knots uniformly spaced for angles between $0$ and $\pi$. The intermolecular CG interactions were represented as the sum of an Akima spline and a WCA repulsion having the form of eq.~\eqref{eq:wca} but with $\sigma$ replaced by $\sigma_{ij}$ for each pair of CG bead types $i$ and $j$. The value of $\sigma_{ij}$ for each pair was manually chosen based on the radial distribution function $g_{ij}^{\rm FG}$ for CG bead types $i$ and $j$ from mapped FG configurations. We used $\sigma_{\rm CC} = 2\,\sigma$, $1.5\,\sigma$, and $1.9\,\sigma$ for dilute, intermediate, and semidilute concentrations of tetra-PEG, respectively, and $\sigma_{\rm CA} = \sigma_{\rm AA} = \sigma$ for the other two pairs for all concentrations. The Akima splines for the dilute concentrations had 12 uniformly spaced knots over distances from $1\,\sigma$ to $4\,\sigma$, whereas for the intermediate and semidilute concentrations, we used a broader range from $0.9\,\sigma$ to $6\,\sigma$ to allow the potentials to adjust both the short-ranged repulsion and longer-ranged interactions. These splines had 20 and 41 uniformly spaced knots for the two concentrations, respectively. The overlap between the two parts of the potential allowed the Akima spline to modify some of the repulsion provided by the WCA potential if necessary. The parameters of the interaction potentials that could be optimized were the differences in value between adjacent knots \cite{Lindquist2016Communication:Optimization,Jadrich2017ProbabilisticMaterials}.

Initial guesses for the parameters of the bond and angle potentials were obtained by hand fitting the mapped FG distributions to potentials with the functional form,
\begin{equation}
f(x) =
\begin{cases}
k_{\rm B} T \left[-\ln(e_0) + e_2(e_1-x)^{e_3} \right], & x \le e_1, \\
k_{\rm B} T \left[-\ln(e_0) + e_4(x-e_1)^{e_5} \right], & x > e_1.
\end{cases}, 
\label{eq:bonded_guess}
\end{equation}
where $x$ is either the bond length or the angle. This functional form is arbitrary, but we found it to describe the data reasonably well. The best-fit coefficients are provided in Table~\ref{tab:bond_angle_params}.
\begin{table}
\caption{Best fit parameters of eq.~\eqref{eq:bonded_guess} for the bond and angle potentials.}
\begin{tabular}{c@{\hspace{1.5em}}c@{\hspace{1.5em}}c}
parameter & bond & angle \\
\hline
$e_0$ & $0.004$ & $0.420$ \\
$e_1$ & $3.050$ & $2.812$ \\
$e_2$ & $0.444$ & $0.026$ \\
$e_3$ & $2.350$ & $4.439$ \\
$e_4$ & $0.656$ & $0.016$ \\
$e_5$ & $2.374$ & $0.178$
\end{tabular}
\label{tab:bond_angle_params}
\end{table}

The parameters of the Akima spline potentials were then fit to eq.~\eqref{eq:bonded_guess}. Initial guesses of the parameters of the pair potentials were generated by Boltzmann inversion \cite{ruhle:macromoltheorysimul:2011} of the radial distribution functions computed from mapped FG configurations using histograms with bin widths of $0.05\,\sigma$ and shifted to zero at their cutoffs. These potentials were manually smoothed or extrapolated to avoid large, unphysical features from statistical uncertainty or unsampled configurations.

The parameters of the interaction potentials were then varied to minimize $S_{\rm rel}$ using the Adam optimizer \cite{adamoptimization}. Specifically, the parameters at iteration $n$, denoted $\boldsymbol{\theta}^{(n)}$, were updated starting from $n=1$ according to
\begin{equation}
    \theta_i^{(n)} = \theta_i^{(n-1)} - \alpha \frac{m_i^{(n)}/(1-\beta_1^n)}{[v_i^{(n)}/(1-\beta_2^n)]^{1/2} + \epsilon},
    \label{eq:adam}
\end{equation}
where $i$ denotes a vector component,
\begin{align}
m_i^{(n)} &= \beta_1 m_i^{(n-1)} + (1-\beta_1) g_i^{(n-1)}, \\
v_i^{(n)} &= \beta_2 v_i^{(n-1)} + (1-\beta_2) (g_i^{(n-1)})^2,
\end{align}
and $\vv{g}^{(n)} = \partial S_{\mathrm{rel}}/\partial \boldsymbol{\theta}$ evaluated at iteration $n$. Here, $\alpha$ is the learning rate, $\beta_1$ and $\beta_2$ control the exponential averaging of the first and second gradient moments, respectively, and $\epsilon$ is a small constant introduced for numerical stability. The learning rates were $\alpha=0.01$ at the dilute concentration and $\alpha = 0.001$ at the intermediate and semidilute concentrations. We used $\beta_1 = 0.9$, $\beta_2 = 0.999$, and $\epsilon=10^{-8}$ for all minimizations, and we initialized the auxiliary variables as $\vv{m}^{(0)} = \vv{0}$ and $\vv{v}^{(0)} = \vv{0}$. Although Adam is widely used in other optimization settings, it has not, to our knowledge, been previously applied to relative-entropy minimization. Here, we found Adam to provide more stable and faster convergence than standard steepest-descent minimization, and unlike Newton--Raphson methods, it required only first derivatives of $S_{\rm rel}$  \cite{shell:jcp:2008, chaimovich:jcp:2011}. We iterated the minimization until good visual agreement between the FG and CG distribution functions was achieved and the potentials were no longer significantly changing. In some cases, isolated spline knots developed sharp kinks in the pair potentials that produced undesired features in the radial distribution functions. These artifacts were removed by replacing the affected knot values with a linear interpolation between neighboring knots. After each correction, a simulation was performed to recompute the distributions and confirm that the corrected potential matched the target distributions.

The minimization procedure was carried out using relentless \cite{Sreenivasan2024Relentless:Optimization} (modified version 0.2.1) with HOOMD-blue \cite{Anderson2020HOOMD-blue:Simulations,HOWARD201645,HOWARD2019139} (version 5.4.0) as a simulation engine. At each iteration, a CG simulation was performed using Langevin dynamics with timestep $0.001\,\tau$ and friction coefficient $0.1\,m/\tau$. The interaction potentials were specified in the simulation as tables with $10^3$ uniformly spaced points. The polymers were equilibrated for $500\,\tau$, then configurations were recorded every $1\,\tau$ for $10^3\,\tau$ for analysis.

At the dilute concentration, we optimized all interaction parameters simultaneously using eq.~\eqref{eq:gradgen} to evaluate $\partial S_{\rm rel}/\partial\boldsymbol{\theta}$, achieving agreement with the FG model after 7 iterations. The same procedure was initially attempted for the intermediate and semidilute concentrations, but these calculations proved to be more difficult. The bond and angle potentials changed little from their Boltzmann-inverted initial guesses at the dilute concentration so, to simplify, we held the bond and angle potentials fixed at their initial guesses and optimized only the intermolecular pair potentials $u_{ij}$ between CG bead types $i$ and $j$ for the intermediate and semidilute concentrations. For these calculations, we computed $\partial S_{\rm rel}/\partial\boldsymbol{\theta}$ using a simplification of eq.~\eqref{eq:gradgen} for pair potentials,
\begin{align}
\frac{\partial S_{\mathrm{rel}}}{\partial \boldsymbol\theta} = &\sum_{i} \sum_{j} \frac{2 \pi \beta N_i N_j}{V} \int_0^{\infty} \left[g_{ij}^{\rm FG}(r)-g_{ij}^{\rm CG}(r ; \boldsymbol\theta)\right] \frac{\partial u_{ij}}{\partial \boldsymbol\theta} r^2 \d{r},
\label{eq:gradS}
\end{align}
where the sums run over all particle types, $N_i$ is the number of CG beads of type $i$, $V$ is the volume, and $g_{ij}^{\rm CG}$ is the radial distribution function for CG bead types $i$ and $j$ from CG configurations. The radial distribution functions were computed using freud (version 3.5.0)~\cite{RAMASUBRAMANI2020107275} up to distance $8\,\sigma$ with bin width $0.05\,\sigma$ and accounting for intramolecular exclusions. The CG models at the intermediate and semidilute concentrations achieved agreement with the FG model after 178 and 313 iterations, respectively. In all cases, we then performed one additional iteration using the general gradient of eq.~\eqref{eq:gradgen} with parameters of both the intramolecular and intermolecular interactions included in $\boldsymbol{\theta}$ to verify the design, confirming that the bond and angle potentials did not change during the update. 
 
\subsection{Dynamic bonding and network formation}
\label{sec:methods:network}
To model network formation due to reversible bonding that occurs between the terminal beads of star arms, we used a hybrid Langevin dynamics/Monte Carlo approach developed by Hocky and co-workers \cite{Mitra2022ABinders}. The same approach and interactions for bonding were adopted for both the FG and CG models. Mimicking experiments where two types of stars are functionalized with complementary dynamic bonding pairs, half of the stars in the simulations were considered type A and the other half type B. Only cross-species reactions between an arm terminus of a type A star and that of a type B star were allowed. During the course of a Langevin dynamics simulation, A--B bonds had the opportunity to form between neighboring stars every $0.01\,\tau$. Bonds formed with probabilities calculated based on the forward and reverse rate constants, $k_{\text{on}}$ and $k_{\text{off}}$~\cite{Mitra2022ABinders}.  We fixed $k_{\text{on}}=100\,\tau^{-1}$ and varied $k_{\text{off}}$ in different simulations to understand the role of the effective A--B bond strength $\varepsilon = \ln(k_{\text{on}}/k_{\text{off}})\,k_{\text{B}}T$. Dynamic bonds were modeled using a harmonic potential with spring constant $k=60\,k_{\text{B}}T/\sigma^2$ and rest length $r_{\text{0}} = 0.98\,\sigma$. Langevin dynamics with a friction coefficient $0.5 \,m/\tau$ was used to simulate the CG tetra-PEG model, and $0.1 \,m/\tau$ for the FG tetra-PEG model. Initially, complementary star polymers, tetra-PEG A and tetra-PEG B, were placed in equal proportion into a cubic box of length $200 \, \sigma$ with periodic boundary conditions, but with bonding turned off. The box was then resized using a volume ramp to scale the box lengths at a constant rate over over $2 \times 10^3\,\tau$ to a lower target density, after which stars were allowed to freely mix until they equilibrated. Bonding was then turned on, and data was collected every $50\,\tau$ during production simulations of $5\times 10^4\,\tau$.

\section{Results \label{results}}
\subsection{Structure}
\begin{figure}
 \includegraphics[width=0.8\textwidth]{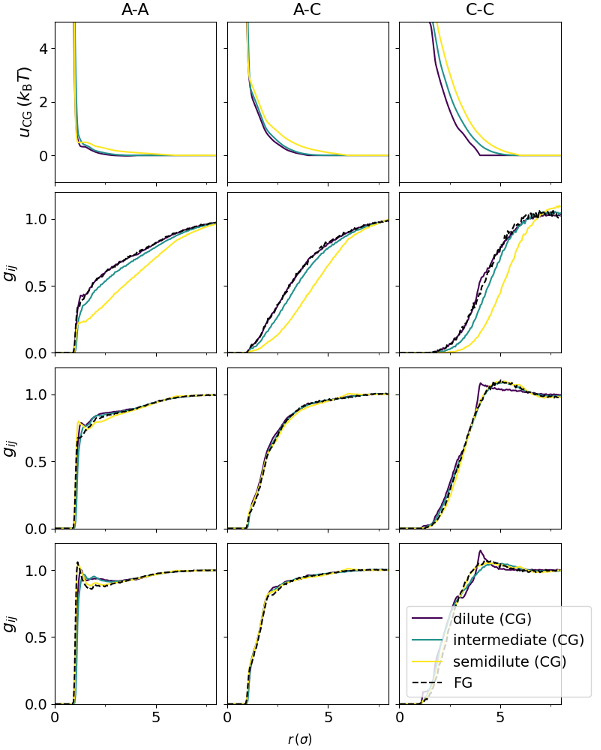}
 \caption{\label{fig:rdfs} {\bf{Transferability of CG model.}} Top row: CG pair potential $u_{\rm CG}$ as a function of distance $r$ for the arm--arm (A--A) (left), arm--center (A--C) (middle), and center--center (C--C) (right) interactions designed at the dilute (purple), intermediate (green), and semidilute (yellow) densities. Radial distribution function $g_{ij}$ calculated using each CG potential at dilute ($\rho = 0.002\,\sigma^{-3}$, second row), intermediate ($\rho = 0.011\,\sigma^{-3}$, third row), and semidilute ($\rho = 0.019\,\sigma^{-3}$, bottom row) conditions. Comparisons highlight that the CG model optimized at intermediate density approximately reproduces the structural correlations of the FG model (dashed line) at all densities.}
 \end{figure}

The first step for validating our systematic coarse-graining approach was to compare the static structural correlations in the FG model to the corresponding CG models parametrized at dilute ($\rho=0.002\,\sigma^{-3}$), intermediate ($\rho=0.011\,\sigma^{-3}$), and semidilute ($\rho=0.019\,\sigma^{-3}$) conditions, in the absence of inter-star linking. To understand the transferability of CG model parameters, structural predictions of CG models optimized at different conditions were compared to the FG model at each state point (Fig.~\ref{fig:rdfs}). The optimized arm--arm (A--A) inter-star pair potentials of the CG model for all three conditions were short-ranged (nearly hard-sphere-like) repulsions, while the repulsive arm--center (A--C) and center--center (C--C) inter-star pair potentials featured significantly longer-ranged repulsions (top row of Fig.~\ref{fig:rdfs}). These longer-ranged A--C and C--C repulsions for the CG model were required to compensate for its less-dense--and thus less intrinsically repulsive--explicit corona compared to the FG model. Accordingly, and as expected, increasingly stronger A--C and C--C repulsions in the CG model were required to match the behavior of the FG model at higher star densities. Despite these quantitative discrepancies in the CG pair potentials optimized under different conditions, parameters optimized at either intermediate or dilute densities enabled the CG model to approximately reproduce the FG structure across all densities studied, from the dilute to semidilute regime.  

\begin{figure}
\includegraphics[width=\textwidth]{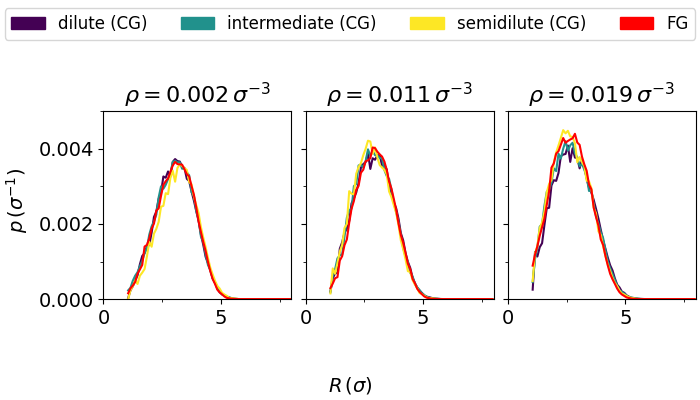}
\caption{\label{fig:e2e} {\bf{Intra-star correlations.}} Probability density $p$ of arm length $R$ in the CG and FG models with a linking strength of $\varepsilon = 11\,k_{\text{B}}T$.}
\end{figure}

We also compared the distribution of the arm length $R$ between the FG and CG models at a linking strength of $\varepsilon = 11\,k_{\text{B}}T$ to understand how inter-star linking might influence intramolecular structural correlations (Fig. \ref{fig:e2e}). For this comparison, as before, the three CG models were simulated at all densities. Notably, there was quantitative agreement between the probability density $p(R)$ for the FG model and the CG model optimized for the density of interest, even after inter-star linking was activated in both models. Interestingly, CG models designed at different conditions also produced distributions of arm lengths that approximately matched the FG model once bonding was enabled. In other words, as far as intramolecular correlations were concerned, the CG interactions between stars could be effectively designed in the absence of linking (i.e., independently of the inter-star bond reaction parameters), without fear of anomalous collapse or swelling once inter-star association and network formation were present.    
 
\begin{figure}
\includegraphics[width=\textwidth]{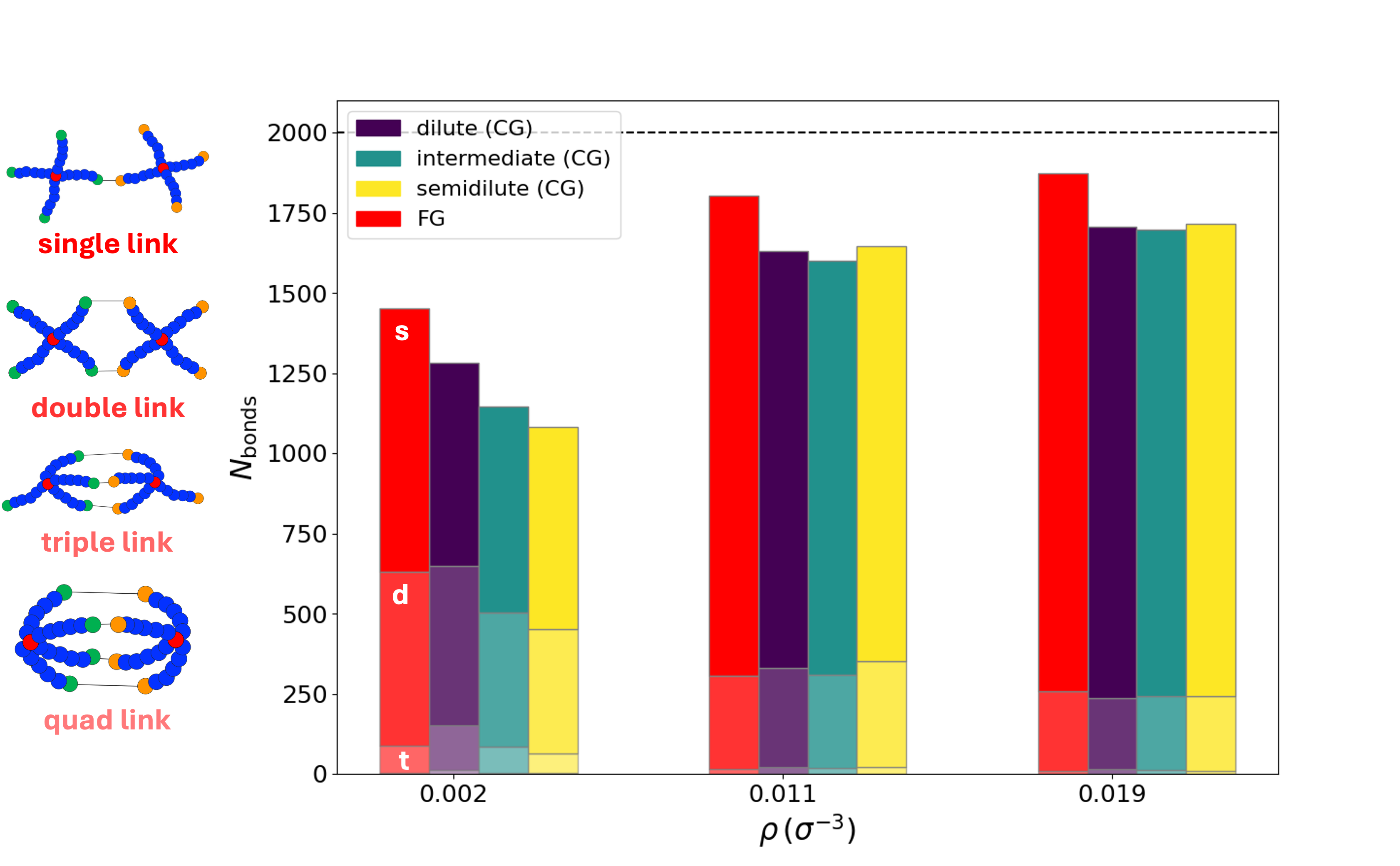}
\caption{{\bf{Inter-star bonding motifs.}} Total number of bonds and number of bonds for different motifs formed in polymer network using CG and FG models. From darkest to lightest shade, bonds were categorized as: single link (s), double link (d), triple link (t), or quad link (q). Quadruple links are present in quantities less than $0.0089 \%$ of total bonds and are therefore not visible on the plot. The dashed line shows the maximum number of possible bonds in the simulated networks. \label{fig:links}}
\end{figure}

We next investigated how linking impacted inter-star network formation. Inter-star bonding motifs were analyzed for the three CG models and compared to those of the FG model at dilute, intermediate, and semidilute densities and $\varepsilon = 11\,k_{\text{B}}T$ (Fig.~\ref{fig:links}). For reference, a fully linked, simulated network would have 2000 inter-star bonds with no defects, based on 2 bonds per polymer and 1000 polymers in the periodically replicated simulation cell. Although a fully linked network was not favored at this $\varepsilon$ for either the CG or FG models, the number of network bonds monotonically increased with density for all models in the same qualitative manner, exceeding 85\% of the maximum bonds at the semidilute state point. Though the number of bonds formed in each of the CG models qualitatively tracked that of the FG model, the CG models consistently formed between 5\% and 15\% fewer bonds than the FG model, with the model parametrized at the semidilute condition showing the largest discrepancy. This highlights one consequence of the more repulsive interactions between individual CG model beads required to reproduce the effective repulsions and inter-star structuring of the reference FG model at higher density. Although this discrepancy in the total number of bonds between the CG and FG models is quantitatively small, the consequences for phase behavior were significant, as discussed further below. 

In addition to the total number of inter-star links formed, the specific star bonding motifs in the CG and FG models were also compared (Fig.~\ref{fig:links}). There are four possible bonding configurations between a star and a neighbor: an ideal (single) bond, a double bond, a triple bond, and a quadruple bond. The double, triple, and quadruple bond motifs were considered defective, as they inhibited extending the ideal network and decreased the average valence, or number of connected neighbors, for stars below the ideal valence number of 4. In general, the fraction of defective bonds formed in the CG models also tracked that of the FG model across the density range studied. Lower defective bonding fractions were observed at higher densities in all models, as expected, though the CG models exhibited a slightly higher defective bonding fraction than the FG model.

In summary, we observed that the structural correlations of the reference FG model at a given density (in the absence of inter-star linking) were well reproduced by the CG model optimized at the same density. Moreover, the CG model optimized at dilute or intermediate densities could adequately reproduce the reference FG model structural correlations at all densities. However, this matching required the CG models to exhibit stronger bead--bead repulsions, which slightly reduced their propensity to form ideal, inter-star links with dynamic bonding present. As detailed next, these seemingly small differences in network formation manifested in significant differences in the phase behaviors predicted by the CG and FG models.

\subsection{Phase behavior}
Having characterized the CG and FG models, we proceeded to explore the phase stability of the star polymers as a function of number density $\rho$ and inter-star bond strength $\varepsilon$. To determine the state of the polymers, the static structure factor of the central beads extrapolated to zero wavevector, $S(0) = \lim_{q \to 0} S(q)$, was estimated from simulated configurations of the CG and FG models. $S(q)$ was computed by averaging bins of width $0.1 \sigma$. The extrapolation was carried out by fitting the low-$q$ structure factor for a range of wavevectors, $q < 5\,\sigma^{-1}$ to $S(q)\approx S(0)/[1+(q\xi)^2]$~\cite{fisher1967theory}. In the thermodynamic limit, $S(0)$ diverges when a single-phase system becomes unstable, while in a finite-size simulation, pronounced increases in $S(0)$ signal a propensity to phase separate~\cite{lindquist2016formation, howard2019structure,howard2021effects}.

\begin{figure}
\includegraphics[width=\textwidth]{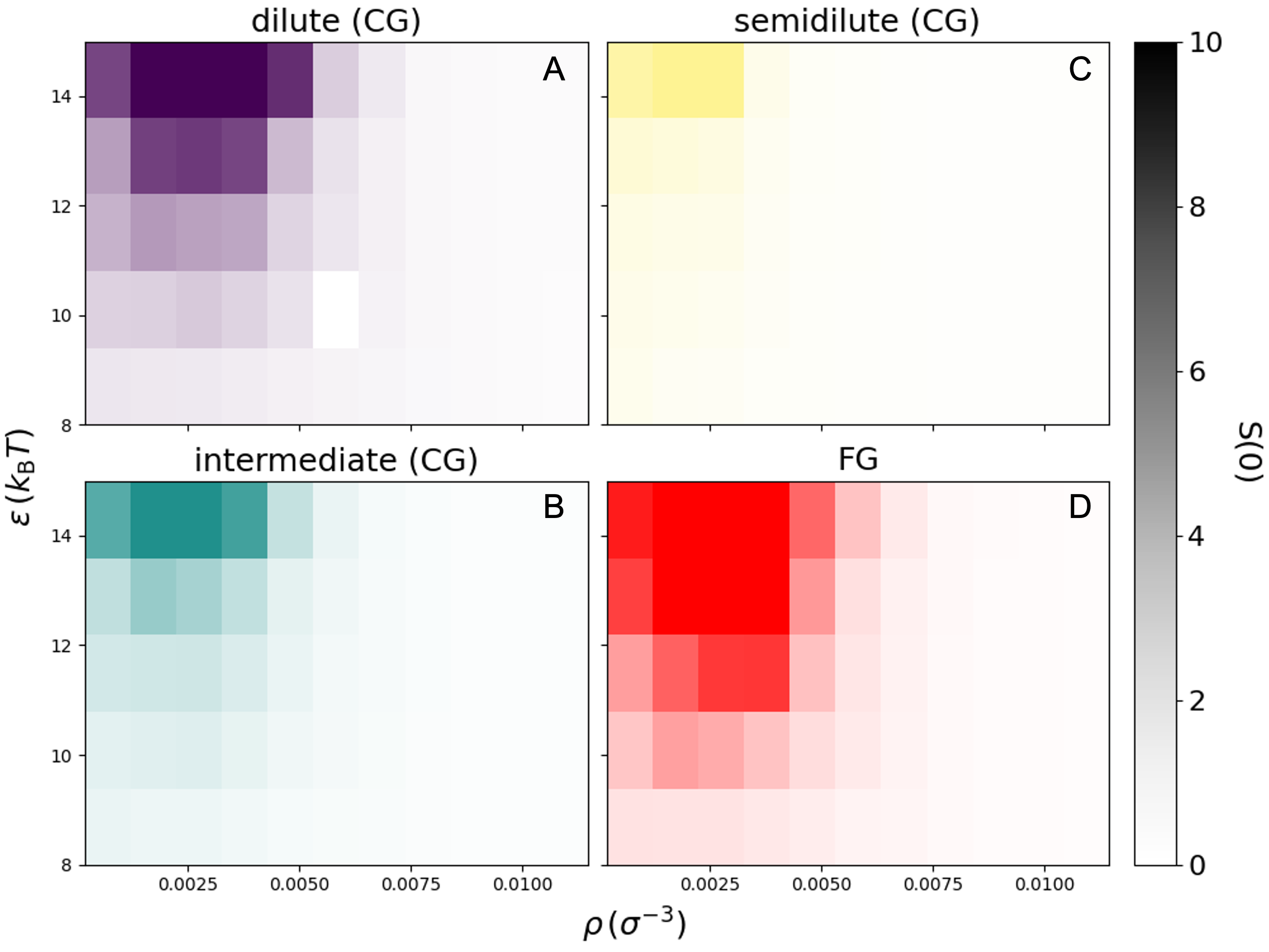}
\caption{{\bf{Phase behavior}}. Comparison of propensity for phase separation, as quantified by the magnitude of the static structure factor extrapolated to zero wavevector $S(0)$ for CG models parametrized at (A) dilute, (B) intermediate, and (C) semidilute conditions as well as (D) the FG model.\label{fig:phase}}
\end{figure}

Based on the structure factor analysis of CG and FG models, macroscopic phase separation into dilute and intermediate-density tetra-PEG phases was predicted once the inter-star bonding strength $\varepsilon$ became sufficiently high relative to the thermal energy scale (Fig.~\ref{fig:phase}). The intermediate-density phase comprised a percolated tetra-PEG network, while the dilute phase consisted of only small star clusters and isolated tetra-PEG stars. While the FG model and the CG model parametrized at the dilute condition both predicted the propensity for phase separation to become strong for bond strengths $\gtrsim 12\,k_{\rm B}T$, phase separation was not anticipated unless bond strengths exceeded $14\,k_{\rm B}T$ for the CG models parametrized at either intermediate or semidilute conditions.    

Because phase separation requires capturing the thermodynamic properties of two coexisting phases, it challenges any CG model parametrized at a single state point. Recall that the CG model parametrized at dilute conditions, with the least repulsive interactions, most accurately captured the FG model's inter-star bonding in both the dilute and intermediate-density phases (Fig.~\ref{fig:links}). The comparisons in Fig.~\ref{fig:phase} suggest that this ability to capture bonding propensity in the dilute and intermediate density phases is key for accurately reproducing the phase behavior of the FG model. 

\subsection{Computational implications of coarse-graining}
\begin{figure}
 \includegraphics[width=10cm]{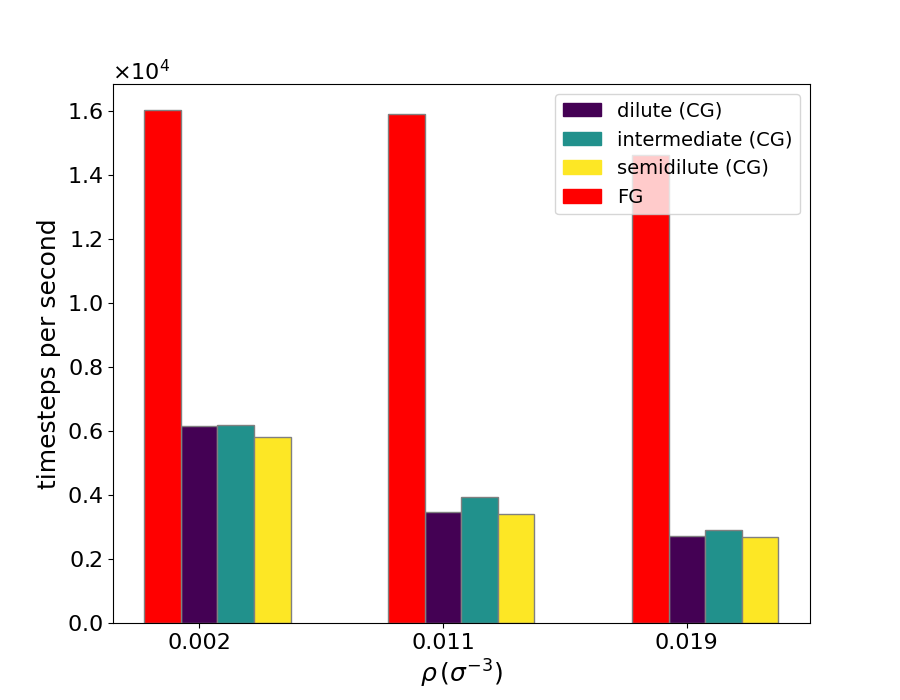}%
 \caption{{\bf{Performance benchmarks for CG and FG models.}} Simulation timesteps per second of a Langevin dynamics simulation for the FG model (red) and CG models parametrized under dilute (purple), intermediate (green), and semidilute (yellow) conditions with no inter-star linking, which is treated the same in the CG and FG models.\label{fig:bench}}
 \end{figure}

The practical objective of using a CG model is typically to decrease the size of the model and/or time required for a simulation, thereby allowing larger and/or longer simulations to be conducted. By construction, the CG model had fewer beads than the FG model and so had a smaller model size. We also conducted benchmark simulations using the CG models and FG model at different densities. These simulations were performed on one NVIDIA A100 GPU on Texas Advanced Computing Center's (TACC) Lonestar cluster. The benchmark simulations underwent a warmup period of $2.5\times 10^3\,\tau$ before collecting time data for a $50\,\tau$ period, which was averaged over 5 runs. Surprisingly, we found that the FG model was consistently faster to simulate than the CG models (Fig.~\ref{fig:bench}).  

Although the CG model had fewer beads than the FG model, its intermolecular pair potentials needed to be significantly longer-ranged to capture the effective interactions in the FG model (Fig.~\ref{fig:rdfs}). Thus, each bead in the CG model had many more pairwise interactions than a bead in the FG model. We expect the simulation timesteps per second to scale roughly inversely with the total number of pairwise interactions, i.e., approximately as $\sim 1/(N_{\rm b}^2 r_{\rm c}^3)$ where $N_{\rm b}$ is the total number of beads in the model and $r_{\rm c}$ is the typical cutoff distance for the pairwise interactions. Since $r_{\rm c}$ is roughly 5 times longer but $N_{\rm b}$ is five times smaller in the CG model than in the FG model, a CG simulation is, undesirably, estimated to be about five times slower than an equivalent FG simulation. The benchmark simulations are reasonably consistent with this estimate.
 
This practical result offers a cautionary note for developing systematically coarse-grained models for star polymers when computational efficiency is an important objective. The CG model should become more computationally favorable as the complexity of the FG model increases, e.g., if an atomistic FG model were used rather than a bead--spring model. Future work might consider CG mapping strategies that strike a favorable balance between reducing the number of interaction sites and increasing the range of the interactions.
 
\section{Conclusions \label{conclusion}}
Here, we introduced a minimalist, 5-bead model to study interactions between reversibly associating tetra-PEG star polymers, which we systematically coarse-grained from and compared with a finer-grained bead--spring model. Our results for the structural correlations, bonding motifs, and phase behavior of tetra-PEG networks show how the accuracy and transferability of CG model parameters depend on the tetra-PEG density used for coarse-graining. They also highlight intrinsic trade-offs between reducing the number of coarse-grained beads and increasing the range of the coarse-grained interactions, a balance that determines the model's computational efficiency. Future implementations of this strategy may demand a more rigorous assessment of how to predict and balance these trade-offs based on the configurational properties of the fine-grained model. 

Looking forward, there is an opportunity to develop more sophisticated, multistate optimization approaches~\cite{moore2014derivation,jones2026msibi,rosenberger2019relative} to create coarse-grained star-polymer models with transferable parameters. We also need to develop more accurate fine-grained star-polymer models based on systematic coarse-graining of atomistic models. The results of the present work suggest opportunities to assess how star functionality, star-arm length and flexibility, and bonding-reaction kinetics affect bonding motifs, heterogeneity, and the dynamic mechanical response of star polymer networks. Ultimately, appropriate coarse-grained models for macromer network formation may also open the door to inverse methods to discover design rules for synthesizing end-linked star polymers and dendrimers with desired properties~\cite{kadulkar2022machine,xie2026learning}.

\begin{acknowledgments}
We thank Dr. Jing Chen from the Molecular Sciences Software Institute for helpful discussions and guidance during the development of the software that enabled the coarse-graining study. We acknowledge the Army Research Office under grant \#W911NF-23-1-0387, the National Science Foundation under grant \#2323482, and the donors of the American Chemical Society Petroleum Research Fund under grant \#66616-DNI9 for support of this research. C.L.P. was supported by a fellowship from The Molecular Sciences Software Institute under National Science Foundation Award grant \#2136142. This work used the Texas Advanced Computing Center (TACC) and Delta at the National Center for Supercomputing Applications through allocation CHM250046 from the Advanced Cyberinfrastructure Coordination Ecosystem: Services \& Support (ACCESS) program \cite{access}, which is supported by National Science Foundation grants \#2138259, \#2138286, \#2138307, \#2137603, and \#2138296. 
\end{acknowledgments}

\bibliography{references}

@article{Sreenivasan2024Relentless:Optimization,
    title = {relentless: Transparent, reproducible molecular dynamics simulations for optimization},
    year = {2024},
    journal = {J. Chem. Phys.},
    author = {Sreenivasan, Adithya N. and  Petix, C. Levi and Sherman, Zachary M. and Howard, Michael P.},
    number = {21},
    month = {12},
    volume = {161},    pages = {212502},
    publisher = {American Institute of Physics},
    url = {/aip/jcp/article/161/21/212502/3323766/relentless-Transparent-reproducible-molecular},
    doi = {10.1063/5.0233683},
    issn = {10897690},
    pmid = {39629765},
    arxivId = {2408.03213}
}

@article{Anderson2020HOOMD-blue:Simulations,
    title = {{HOOMD-blue}: {A Python} package for high-performance molecular dynamics and hard particle {Monte Carlo} simulations},
    year = {2020},
    journal = {Computational Materials Science},
    author = {Anderson, Joshua A. and Glaser, Jens and Glotzer, Sharon C.},
    month = {2},
    pages = {109363},
    volume = {173},
    doi = {10.1016/j.commatsci.2019.109363},
    issn = {09270256}
}

@article{Bianchi2015Soft-patchySelf-organization,
    title = {Soft-patchy nanoparticles: modeling and self-organization},
    year = {2015},
    journal = {Faraday Discussions},
    author = {Bianchi, Emanuela and Capone, Barbara and Kahl, Gerhard and Likos, Christos N.},
    number = {0},
    month = {7},
    pages = {123--138},
    volume = {181},
    publisher = {The Royal Society of Chemistry},
    url = {https://pubs.rsc.org/en/content/articlehtml/2015/fd/c4fd00271g https://pubs.rsc.org/en/content/articlelanding/2015/fd/c4fd00271g},
    doi = {10.1039/c4fd00271g},
    issn = {13645498}
}

@article{Bertsch2023Self-HealingRegeneration,
    title = {Self-Healing Injectable Hydrogels for Tissue Regeneration},
    year = {2023},
    journal = {Chem. Rev.},
    author = {Bertsch, Pascal and Diba, Mani and Mooney, David J. and Leeuwenburgh, Sander C.G.},
    number = {2},
    month = {1},
    pages = {834--873},
    volume = {123},
    publisher = {American Chemical Society},
    doi = {10.1021/acs.chemrev.2c00179},
    issn = {15206890},
    pmid = {35930422}
}

@article{Yesilyurt2016InjectableProperties,
    title = {Injectable Self-Healing Glucose-Responsive Hydrogels with {pH}-Regulated Mechanical Properties},
    year = {2016},
    journal = {Adv. Mater.},
    author = {Yesilyurt, Volkan and Webber, Matthew J. and Appel, Eric A. and Godwin, Colin and Langer, Robert and Anderson, Daniel G.},
    number = {1},
    month = {1},
    pages = {86--91},
    volume = {28},
    publisher = {Wiley-VCH Verlag},
    doi = {10.1002/adma.201502902},
    issn = {15214095},
    pmid = {26540021}
}

@article{Shibayama2019PrecisionFuture,
  title={Precision polymer network science with tetra-{PEG} gels—a decade history and future},
  author={Shibayama, Mitsuhiro and Li, Xiang and Sakai, Takamasa},
  journal={Colloid Polym. Sci.},
  volume={297},
  number={1},
  pages={1--12},
  year={2019},
  publisher={Springer}
}

@article{Parada2018IdealNetworks,
    title = {Ideal reversible polymer networks},
    year = {2018},
    journal = {Soft Matter},
    author = {Parada, German Alberto and Zhao, Xuanhe},
    number = {25},
    pages = {5186--5196},
    volume = {14},
    publisher = {Royal Society of Chemistry},
    doi = {10.1039/c8sm00646f},
    issn = {17446848},
    pmid = {29780993}
}

@article{Sakai2008DesignMacromonomers,
    title = {Design and Fabrication of a High-Strength Hydrogel with Ideally Homogeneous Network Structure from Tetrahedron-like Macromonomers},
    year = {2008},
    journal = {Macromolecules},
    author = {Sakai, Takamasa and Matsunaga, Takuro and Yamamoto, Yuji and Ito, Chika and Yoshida, Ryo and Suzuki, Shigeki and Sasaki, Nobuo and Shibayama, Mitsuhiro and Chung, Ung Il},
    number = {14},
    month = {7},
    pages = {5379--5384},
    volume = {41},
    publisher = {American Chemical Society},
    url = {/doi/pdf/10.1021/ma800476x?ref=article_openPDF},
    doi = {10.1021/ma800476x},
    issn = {00249297}
}

@article{Fitzsimons2020PreferentialAdditions,
    title = {Preferential Control of Forward Reaction Kinetics in Hydrogels Crosslinked with Reversible Conjugate Additions},
    year = {2020},
    journal = {Macromolecules},
    author = {Fitzsimons, Thomas M. and Oentoro, Felicia and Shanbhag, Tej V. and Anslyn, Eric V. and Rosales, Adrianne M.},
    number = {10},
    month = {5},
    pages = {3738--3746},
    volume = {53},
    publisher = {American Chemical Society},
    doi = {10.1021/acs.macromol.0c00335},
    issn = {15205835}
}

@article{FitzSimons2022EffectBonds,
    title = {Effect of pH on the Properties of Hydrogels Cross-Linked via Dynamic thia-{Michael} Addition Bonds},
    year = {2022},
    journal = {ACS Polymers Au},
    author = {FitzSimons, Thomas M. and Anslyn, Eric V. and Rosales, Adrianne M.},
    number = {2},
    month = {4},
    pages = {129--136},
    volume = {2},
    doi = {10.1021/acspolymersau.1c00049},
    issn = {2694-2453}
}

@article{Apostolides2017DynamicMaterials,
    title = {Dynamic Covalent Star Poly(ethylene glycol) Model Hydrogels: A New Platform for Mechanically Robust, Multifunctional Materials},
    year = {2017},
    journal = {Macromolecules},
    author = {Apostolides, Demetris E. and Sakai, Takamasa and Patrickios, Costas S.},
    number = {5},
    month = {3},
    pages = {2155--2164},
    volume = {50},
    publisher = {American Chemical Society},
    doi = {10.1021/acs.macromol.7b00236},
    issn = {15205835}
}

@article{Yesilyurt2017MixedNetworks,
  title={Mixed reversible covalent crosslink kinetics enable precise, hierarchical mechanical tuning of hydrogel networks},
  author={Yesilyurt, Volkan and Ayoob, Andrew M and Appel, Eric A and Borenstein, Jeffrey T and Langer, Robert and Anderson, Daniel G},
  journal={Adv. Mater.},
  volume={29},
  number={19},
  pages={1605947},
  year={2017},
  publisher={Wiley Online Library}
}

@article{Ahmadi2020DynamicCrosslinks,
    title = {Dynamic Model Metallo-Supramolecular Dual-Network Hydrogels with Independently Tunable Network Crosslinks},
    year = {2020},
    journal = {J. Polym. Sci},
    author = {Ahmadi, Mostafa and Seiffert, Sebastian},
    pages = {330--342},
    volume = {2020},
    url = {https://onlinelibrary.wiley.com/doi/10.1002/pol.20190076},
    doi = {10.1002/pola.20190076}
}

@article{Ahmadi2021CoordinationNetworks,
    title = {Coordination Geometry Preference Regulates the Structure and Dynamics of Metallo-Supramolecular Polymer Networks},
    year = {2021},
    journal = {Macromolecules},
    author = {Ahmadi, Mostafa and Seiffert, Sebastian},
    number = {3},
    month = {2},
    pages = {1388--1400},
    volume = {54},
    doi = {10.1021/acs.macromol.0c02524},
    issn = {0024-9297}
}

@article{Ahmadi2020ThermodynamicCoordination,
    title = {Thermodynamic control over energy dissipation modes in dual-network hydrogels based on metal-ligand coordination},
    year = {2020},
    journal = {Soft Matter},
    author = {Ahmadi, Mostafa and Seiffert, Sebastian},
    number = {9},
    month = {3},
    pages = {2332--2341},
    volume = {16},
    publisher = {Royal Society of Chemistry},
    doi = {10.1039/c9sm02149c},
    issn = {17446848},
    pmid = {32053126}
}

@article{Shibayama2018Gels:BioMatter,
    title = {Gels: From Soft Matter to BioMatter},
    year = {2018},
    journal = {Ind. Eng. Chem. Res.},
    author = {Shibayama, Mitsuhiro and Li, Xiang and Sakai, Takamasa},
    number = {4},
    month = {1},
    pages = {1121--1128},
    volume = {57},
    doi = {10.1021/acs.iecr.7b04614},
    issn = {0888-5885}
}

@article{Rovigatti2014AccurateNanostars,
    title = {Accurate phase diagram of tetravalent {DNA} nanostars},
    year = {2014},
    journal = {J. Chem. Phys.},
pages = {154903},
    
    author = {Rovigatti, Lorenzo and Bomboi, Francesca and Sciortino, Francesco},
    number = {15},
    month = {4},
    volume = {140}
}

@article{reinhardt2013computing,
  title={Computing phase diagrams for a quasicrystal-forming patchy-particle system},
  author={Reinhardt, Aleks and Romano, Flavio and Doye, Jonathan PK},
  journal={Phys. Rev. Lett.},
  volume={110},
  number={25},
  pages={255503},
  year={2013},
  publisher={APS}
}

@article{liu2016diamond,
  title={Diamond family of nanoparticle superlattices},
  author={Liu, Wenyan and Tagawa, Miho and Xin, Huolin L and Wang, Tong and Emamy, Hamed and Li, Huilin and Yager, Kevin G and Starr, Francis W and Tkachenko, Alexei V and Gang, Oleg},
  journal={Science},
  volume={351},
  number={6273},
  pages={582--586},
  year={2016},
  publisher={American Association for the Advancement of Science}
}

@article{zhang2005self,
  title={Self-assembly of patchy particles into diamond structures through molecular mimicry},
  author={Zhang and Keys, Aaron S and Chen, Ting and Glotzer, Sharon C},
  journal={Langmuir},
  volume={21},
  number={25},
  pages={11547--11551},
  year={2005},
  publisher={ACS Publications}
}

@article{lindquist2016formation,
  title={On the formation of equilibrium gels via a macroscopic bond limitation},
  author={Lindquist, Beth A and Jadrich, Ryan B and Milliron, Delia J and Truskett, Thomas M},
  journal={J. Chem. Phys.},
  volume={145},
  number={7},
  year={2016},
  pages={074906},
  publisher={AIP Publishing}
}

@article{van2012formation,
  title={Formation of dodecagonal quasicrystals in two-dimensional systems of patchy particles},
  author={van der Linden, Marjolein N and Doye, Jonathan PK and Louis, Ard A},
  journal={J. Chem. Phys.},
  volume={136},
  number={5},
  year={2012},
  pages={054904},
  publisher={AIP Publishing}
}

@article{conrad2022emulsion,
  title={Emulsion imaging of a {DNA} nanostar condensate phase diagram reveals valence and electrostatic effects},
  author={Conrad, Nathaniel and Chang, Grace and Fygenson, Deborah K and Saleh, Omar A},
  journal={J. Chem. Phys.},
  volume={157},
  number={23},
  year={2022},
  pages={234203},
  publisher={AIP Publishing}
}

@article{biffi2015equilibrium,
  title={Equilibrium gels of low-valence {DNA} nanostars: a colloidal model for strong glass formers},
  author={Biffi, Silvia and Cerbino, Roberto and Nava, Giovanni and Bomboi, Francesca and Sciortino, Francesco and Bellini, Tommaso},
  journal={Soft Matter},
  volume={11},
  number={16},
  pages={3132--3138},
  year={2015},
  publisher={The Royal Society of Chemistry}
}

@article{Lattuada2021SpatiallyGels,
  title={Spatially uniform dynamics in equilibrium colloidal gels},
  author={Lattuada, Enrico and Caprara, Debora and Piazza, Roberto and Sciortino, Francesco},
  journal={Sci. Adv.},
  volume={7},
  number={49},
  pages={eabk2360},
  year={2021},
  publisher={American Association for the Advancement of Science}
}

@article{Bianchi2006PhaseLiquids,
    title = {Phase Diagram of Patchy Colloids: Towards Empty Liquids},
    year = {2006},
    journal = {Phys. Rev. Lett.},
    author = {Bianchi, Emanuela and Largo, Julio and Tartaglia, Piero and Zaccarelli, Emanuela and Sciortino, Francesco},
    number = {16},
    month = {10},
    pages = {168301},
    volume = {97},
    doi = {10.1103/PhysRevLett.97.168301},
    issn = {0031-9007}
}

@article{Mitra2022ABinders,
  title={A coarse-grained simulation model for colloidal self-assembly via explicit mobile binders},
  author={Mitra, Gaurav and Chang, Chuan and McMullen, Angus and Puchall, Daniela and Brujic, Jasna and Hocky, Glen M},
  journal={Soft Matter},
  volume={19},
  number={23},
  pages={4223--4236},
  year={2023},
  publisher={The Royal Society of Chemistry}
}

@article{Shell2016COARSEGRAININGENTROPY,
  title={Coarse-Graining with the Relative Entropy},
  author={Shell, M Scott},
  journal={Adv. Chem. Phys.},
  volume={161},
  pages={395--441},
  year={2016},
  publisher={Wiley Online Library}
}

@article{Lindquist2016Communication:Optimization,
    title = {Communication: Inverse design for self-assembly via on-the-fly optimization},
    year = {2016},
    journal = {J. Chem. Phys.},
    author = {Lindquist, Beth A. and Jadrich, Ryan B. and Truskett, Thomas M.},
    number = {11},
    month = {9},
    volume = {145},
    pages={111101},
    doi = {10.1063/1.4962754},
    issn = {0021-9606}
}

@article{Jadrich2017ProbabilisticMaterials,
    title = {Probabilistic inverse design for self-assembling materials},
    year = {2017},
    journal = {J. Chem. Phys.},
    author = {Jadrich, R. B. and Lindquist, B. A. and Truskett, T. M.},
    number = {18},
    month = {5},
    volume = {146},
    pages={184103},
    doi = {10.1063/1.4981796},
    issn = {0021-9606}
}

@article{Holoman2025SimulatingMaterials,
  title={Simulating dynamic bonding in soft materials},
  author={Holoman, Tyla R and Prajwal, BP and Hocky, Glen M and Truskett, Thomas M},
  journal={Curr. Opin. Colloid Interface Sci.},
  volume={83},
  pages={102019},
  year={2026},
  publisher={Elsevier}
}

@article{Kremer1990DynamicsSimulation,
    title = {Dynamics of entangled linear polymer melts: A molecular-dynamics simulation},
    year = {1990},
    journal = {J. Chem. Phys.},
    author = {Kremer, Kurt and Grest, Gary S.},
    number = {8},
    pages = {5057--5086},
    volume = {92},
    doi = {10.1063/1.458541},
    issn = {00219606}
}

@article{Furuya2020MolecularProperties,
    title = {Molecular simulation of networks formed by end-linking of tetra-arm star polymers: Effects of network structures on mechanical properties},
    year = {2020},
    journal = {Polymer},
    author = {Furuya, Tsutomu and Koga, Tsuyoshi},
    month = {2},
    volume = {189},
    pages= {122195},
    publisher = {Elsevier Ltd}
}

@inproceedings{access,
author = {Boerner, Timothy J. and Deems, Stephen and Furlani, Thomas R. and Knuth, Shelley L. and Towns, John},
title = {ACCESS: Advancing Innovation: NSF’s Advanced Cyberinfrastructure Coordination Ecosystem: Services \& Support},
year = {2023},
isbn = {9781450399852},
publisher = {Association for Computing Machinery},
address = {New York, NY, USA},
doi = {10.1145/3569951.3597559},
booktitle = {Practice and Experience in Advanced Research Computing 2023: Computing for the Common Good},
pages = {173–176},
numpages = {4},
location = {Portland, OR, USA},
series = {PEARC '23}
}

@article{shell:jcp:2008,
	title        = {The relative entropy is fundamental to multiscale and inverse thermodynamic problems},
	author       = {Shell, M. Scott},
	year         = 2008,
	month        = {Oct},
	journal      = {J. Chem. Phys.},
	volume       = 129,
	number       = 14,
	pages        = 144108,
	doi          = {10.1063/1.2992060}
}

@article{weeks:jcp:1971, title={Role of Repulsive Forces in Determining the Equilibrium Structure of Simple Liquids}, volume={54}, DOI={10.1063/1.1674820}, number={12}, journal={J. Chem. Phys.}, author={Weeks, John D. and Chandler, David and Andersen, Hans C.}, year={1971}, month={Jun}, pages={5237–5247} }

@article{ruhle:macromoltheorysimul:2011,
  title={Hybrid approaches to coarse-graining using the {VOTCA} package: liquid hexane},
  author={R{\"u}hle, Victor and Junghans, Christoph},
  journal={Macromol. Theory Simul.},
  volume={20},
  number={7},
  pages={472--477},
  year={2011},
  publisher={Wiley Online Library}
}

@article{adamoptimization,
	title        = {Adam: A Method for Stochastic Optimization},
	author       = {Kingma, Diederik P. and Ba, Jimmy},
	year         = 2017,
	month        = {Jan},
    journal      = {arXiv},
	number       = {arXiv:1412.6980},
	doi          = {10.48550/arXiv.1412.6980}
}

@article{chaimovich:jcp:2011,
	title        = {Coarse-graining errors and numerical optimization using a relative entropy framework},
	author       = {Chaimovich, Aviel and Shell, M. Scott},
	year         = 2011,
	month        = {Mar},
	journal      = {J. Chem. Phys.},
	volume       = 134,
	number       = 9,
	pages        = {094112},
	doi          = {10.1063/1.3557038}
}

@book{Rubenstein2003PolymerPhysics,
    title = {Polymer Physics},
    year = {2003},
    author = {Rubenstein, Michael and Colby, Ralph H.},
    month = {6},
    publisher = {Oxford University Press}
}

@article{Asai2013CorrelationGels,
    title = {Correlation between Local and Global Inhomogeneities of Chemical Gels},
    year = {2013},
    journal = {Macromolecules},
    author = {Asai, Makoto and Katashima, Takuya and Chung, Ung-il and Sakai, Takamasa and Shibayama, Mitsuhiro},
    number = {24},
    month = {12},
    pages = {9772--9781},
    volume = {46},
    doi = {10.1021/ma400486h},
    issn = {0024-9297}
}

@article{green2022,
  title = {Assembling Inorganic Nanocrystal Gels},
  author = {Green, Allison M. and Ofosu, Charles K. and Kang, Jiho and Anslyn, Eric V. and Truskett, Thomas M. and Milliron, Delia J.},
  year = 2022,
  journal = {Nano Lett.},
  volume = {22},
  number = {4},
  pages = {1457-1466--1457-1466},
  doi = {10.1021/acs.nanolett.1c04707}
}

@article{zaccarelli2007colloidal,
  title={Colloidal gels: equilibrium and non-equilibrium routes},
  author={Zaccarelli, Emanuela},
  journal={J. Phys. Condens. Matter},
  volume={19},
  number={32},
  pages={323101},
  year={2007},
  publisher={IOP Publishing}
}

@article{webber2022dynamic,
  title={Dynamic and reconfigurable materials from reversible network interactions},
  author={Webber, Matthew J and Tibbitt, Mark W},
  journal={Nat. Rev. Mater.},
  volume={7},
  number={7},
  pages={541--556},
  year={2022},
  publisher={Nature Publishing Group UK London}
}

@article{tang2021dynamic,
  title={Dynamic covalent hydrogels as biomaterials to mimic the viscoelasticity of soft tissues},
  author={Tang, Shengchang and Richardson, Benjamin M and Anseth, Kristi S},
  journal={Prog. Mater. Sci.},
  volume={120},
  pages={100738},
  year={2021},
  publisher={Elsevier}
}

@article{lee2025dynamic,
  title={Dynamic bond chemistry in soft materials: bridging adaptability and mechanical robustness},
  author={Lee, Haeseung and Kim, Jiyun and Lee, Minwoo and Kang, Jiheong},
  journal={Chem. Rev.},
  volume={125},
  number={23},
  pages={11379--11425},
  year={2025},
  publisher={ACS Publications}
}

@incollection{petekidis2021,
  title = {Rheology of Colloidal Glasses and Gels},
  booktitle = {Theory and {Applications} of {Colloidal} {Suspension} {Rheology}},
  author = {Petekidis, G. and Wagner, N. J.},
  editor = {Wagner, N. J and Mewis, J.},
  year = 2021,
  volume = {8},
  pages = {173--226},
  publisher = {Cambridge University Press},
  address = {Cambridge, UK},
}

@article{wang2015adaptable,
  title={Adaptable hydrogel networks with reversible linkages for tissue engineering},
  author={Wang, Huiyuan and Heilshorn, Sarah C},
  journal={Adv. Mater.},
  volume={27},
  number={25},
  pages={3717--3736},
  year={2015},
  publisher={Wiley Online Library}
}

@article{ollier2023biomimetic,
  title={Biomimetic strain-stiffening in fully synthetic dynamic-covalent hydrogel networks},
  author={Ollier, Rachel C and Xiang, Yuanhui and Yacovelli, Adriana M and Webber, Matthew J},
  journal={Chem. Sci.},
  volume={14},
  number={18},
  pages={4796--4805},
  year={2023},
  publisher={The Royal Society of Chemistry}
}

@article{zheng2024real,
  title={Real-time quantification of molecular-level dynamic behaviors underpinning shear thinning in end-linked associative polymer networks},
  author={Zheng, Yu and Sen, Devosmita and Zou, Weizhong and Dai, Kexin and Olsen, Bradley D},
  journal={J. Am. Chem. Soc.},
  volume={146},
  number={51},
  pages={35285--35294},
  year={2024},
  publisher={ACS Publications}
}

@article{sing2015celebrating,
  title={Celebrating {Soft Matter's} 10$^{\text{th}}$ Anniversary: Chain configuration and rate-dependent mechanical properties in transient networks},
  author={Sing, Michelle K and Wang, Zhen-Gang and McKinley, Gareth H and Olsen, Bradley D},
  journal={Soft Matter},
  volume={11},
  number={11},
  pages={2085--2096},
  year={2015},
  publisher={The Royal Society of Chemistry}
}

@article{mahmad2020understanding,
  title={Understanding the molecular origin of shear thinning in associative polymers through quantification of bond dissociation under shear},
  author={Mahmad Rasid, Irina and Ramirez, Jorge and Olsen, Bradley D and Holten-Andersen, Niels},
  journal={Phys. Rev. Mater.},
  volume={4},
  number={5},
  pages={055602},
  year={2020},
  publisher={APS}
}

@article{crowell2023shear,
  title={Shear thickening behavior in injectable tetra-{PEG} hydrogels cross-linked via dynamic thia-michael addition bonds},
  author={Crowell, Anne D and FitzSimons, Thomas M and Anslyn, Eric V and Schultz, Kelly M and Rosales, Adrianne M},
  journal={Macromolecules},
  volume={56},
  number={19},
  pages={7795},
  year={2023}
}

@article{crowell2026leveraging,
  title={Leveraging bond dissociation kinetics to tune shear-thickening behavior in dynamic covalent tetra-{PEG} hydrogels},
  author={Crowell, Anne D and Kang, Jiho and Conrad, Diana L and FitzSimons, Thomas M and Anslyn, Eric V and Milliron, Delia J and Rosales, Adrianne M},
  journal={Sci. Adv.},
  volume={12},
  number={10},
  pages={eadz9563},
  year={2026},
  publisher={American Association for the Advancement of Science}
}

@article{shibayama1998spatial,
  title={Spatial inhomogeneity and dynamic fluctuations of polymer gels},
  author={Shibayama, Mitsuhiro},
  journal={Macromol. Chem. Phys.},
  volume={199},
  number={1},
  pages={1--30},
  year={1998},
  publisher={Wiley Online Library}
}

@article{miotti2026mesoscopic,
  title={Mesoscopic Modeling of Dynamic Tetra-{PEG} Hydrogel Networks},
  author={Miotti, Pietro and Cousin, Lucien and Tibbitt, Mark W and Pivkin, Igor V},
  journal={arXiv preprint arXiv:2603.17180},
  year={2026}
}

@article{vigil2025coherent,
  title={Coherent state field theory for reversible gelation of star polymers},
  author={Vigil, Daniel L and Fredrickson, Glenn H and Qin, Jian},
  journal={Macromolecules},
  volume={58},
  number={19},
  pages={10920--10936},
  year={2025},
  publisher={ACS Publications}
}

@article{raffaelli2021stress,
  title={Stress relaxation in tunable gels},
  author={Raffaelli, Chiara and Ellenbroek, Wouter G},
  journal={Soft Matter},
  volume={17},
  number={45},
  pages={10254--10262},
  year={2021},
  publisher={The Royal Society of Chemistry}
}

@article{matsunaga2009structure,
  title={Structure characterization of tetra-{PEG} gel by small-angle neutron scattering},
  author={Matsunaga, Takuro and Sakai, Takamasa and Akagi, Yuki and Chung, Ung-il and Shibayama, Mitsuhiro},
  journal={Macromolecules},
  volume={42},
  number={4},
  pages={1344--1351},
  year={2009},
  publisher={ACS Publications}
}

@article{matsunaga2009sans,
  title={{SANS} and {SLS} studies on tetra-arm {PEG} gels in as-prepared and swollen states},
  author={Matsunaga, Takuro and Sakai, Takamasa and Akagi, Yuki and Chung, Ung-il and Shibayama, Mitsuhiro},
  journal={Macromolecules},
  volume={42},
  number={16},
  pages={6245--6252},
  year={2009},
  publisher={ACS Publications}
}

@article{akagi2011examination,
  title={Examination of the theories of rubber elasticity using an ideal polymer network},
  author={Akagi, Yuki and Katashima, Takuya and Katsumoto, Yukiteru and Fujii, Kenta and Matsunaga, Takuro and Chung, Ung-il and Shibayama, Mitsuhiro and Sakai, Takamasa},
  journal={Macromolecules},
  volume={44},
  number={14},
  pages={5817--5821},
  year={2011},
  publisher={ACS Publications}
}

@article{kumar2023reversible,
  title={Reversible networks made of star polymers: Mean-field treatment with consideration of finite loops},
  author={Kumar, Kiran Suresh and Lang, Michael},
  journal={Macromolecules},
  volume={56},
  number={17},
  pages={7166--7183},
  year={2023},
  publisher={ACS Publications}
}

@article{sciortino2011reversible,
  title={Reversible gels of patchy particles},
  author={Sciortino, Francesco and Zaccarelli, Emanuela},
  journal={Curr. Opin. Solid State Mater. Sci.},
  volume={15},
  number={6},
  pages={246--253},
  year={2011},
  publisher={Elsevier}
}

@article{russo2022physics,
  title={The physics of empty liquids: From patchy particles to water},
  author={Russo, John and Leoni, Fabio and Martelli, Fausto and Sciortino, Francesco},
  journal={Rep. Prog. Phys.},
  volume={85},
  number={1},
  pages={016601},
  year={2022},
  publisher={IOP Publishing}
}

@article{rovigatti2014gels,
  title={Gels of {DNA} nanostars never crystallize},
  author={Rovigatti, Lorenzo and Smallenburg, Frank and Romano, Flavio and Sciortino, Francesco},
  journal={ACS Nano},
  volume={8},
  number={4},
  pages={3567--3574},
  year={2014},
  publisher={ACS Publications}
}

@article{smallenburg2013liquids,
  title={Liquids more stable than crystals in particles with limited valence and flexible bonds},
  author={Smallenburg, Frank and Sciortino, Francesco},
  journal={Nat. Phys.},
  volume={9},
  number={9},
  pages={554--558},
  year={2013},
  publisher={Nature Publishing Group UK London}
}

@article{howard2021effects,
  title={Effects of linker flexibility on phase behavior and structure of linked colloidal gels},
  author={Howard, Michael P and Sherman, Zachary M and Sreenivasan, Adithya N and Valenzuela, Stephanie A and Anslyn, Eric V and Milliron, Delia J and Truskett, Thomas M},
  journal={J. Chem. Phys.},
  volume={154},
  number={7},
  year={2021},
  pages={074901},
  publisher={AIP Publishing}
}

@article{howard2019structure,
  title={Structure and phase behavior of polymer-linked colloidal gels},
  author={Howard, Michael P and Jadrich, Ryan B and Lindquist, Beth A and Khabaz, Fardin and Bonnecaze, Roger T and Milliron, Delia J and Truskett, Thomas M},
  journal={J. Chem. Phys.},
  volume={151},
  number={12},
  year={2019},
  pages={124901},
  publisher={AIP Publishing}
}

@article{fisher1967theory,
  title={The theory of equilibrium critical phenomena},
  author={Fisher, Michael E},
  journal={Rep. Prog. Phys.},
  volume={30},
  number={2},
  pages={615--730},
  year={1967}
}

@article{moore2014derivation,
  title={Derivation of coarse-grained potentials via multistate iterative Boltzmann inversion},
  author={Moore, Timothy C and Iacovella, Christopher R and McCabe, Clare},
  journal={J. Chem. Phys.},
  volume={140},
  number={22},
  year={2014},
  pages={224104},
  publisher={AIP Publishing}
}

@article{jones2026msibi,
  title={msibi: Multistate Iterative Boltzmann Inversion},
  author={Jones, Christopher and Almarashi, Mazin and Albooyeh, Marjan and Jankowski, Eric and McCabe, Clare},
  journal={J. Open Source Softw.},
  volume={11},
  number={119},
  year={2026},
  publisher={The Open Journal}
}

@article{rosenberger2019relative,
  title={Relative entropy indicates an ideal concentration for structure-based coarse graining of binary mixtures},
  author={Rosenberger, David and van der Vegt, Nico FA},
  journal={Phys. Rev. E},
  volume={99},
  number={5},
  pages={053308},
  year={2019},
  publisher={APS}
}

@article{kadulkar2022machine,
  title={Machine learning--assisted design of material properties},
  author={Kadulkar, Sanket and Sherman, Zachary M and Ganesan, Venkat and Truskett, Thomas M},
  journal={Ann. Rev. Chem. and Biomol. Eng.},
  volume={13},
  number={2022},
  pages={235--254},
  year={2022},
  publisher={Annual Reviews}
}

@article{xie2026learning,
  title={Learning nature’s assembly language with polymers},
  author={Xie, Oliver and Cohen, Alexander E and Bazant, Martin Z and Olsen, Bradley D},
  journal={Proc. Natl. Acad. Sci. U.S.A.},
  volume={123},
  number={7},
  pages={e2519094123},
  year={2026},
  publisher={National Academy of Sciences}
}

@article{HOWARD201645,
title = {Efficient neighbor list calculation for molecular simulation of colloidal systems using graphics processing units},
journal = {Comput. Phys. Commun.},
volume = {203},
pages = {45-52},
year = {2016},
issn = {0010-4655},
doi = {https://doi.org/10.1016/j.cpc.2016.02.003},
url = {https://www.sciencedirect.com/science/article/pii/S0010465516300182},
author = {Michael P. Howard and Joshua A. Anderson and Arash Nikoubashman and Sharon C. Glotzer and Athanassios Z. Panagiotopoulos}
}

@article{HOWARD2019139,
title = {Quantized bounding volume hierarchies for neighbor search in molecular simulations on graphics processing units},
journal = {Comput. Mater. Sci.},
volume = {164},
pages = {139-146},
year = {2019},
issn = {0927-0256},
doi = {https://doi.org/10.1016/j.commatsci.2019.04.004},
url = {https://www.sciencedirect.com/science/article/pii/S092702561930206X},
author = {Michael P. Howard and Antonia Statt and Felix Madutsa and Thomas M. Truskett and Athanassios Z. Panagiotopoulos}
}

@article{RAMASUBRAMANI2020107275,
title = {freud: A software suite for high throughput analysis of particle simulation data},
journal = {Comput. Phys. Commun.},
author = {Vyas Ramasubramani and Bradley D. Dice and Eric S. Harper and Matthew P. Spellings and Joshua A. Anderson and Sharon C. Glotzer},
volume = {254},
pages = {107275},
year = {2020},
issn = {0010-4655},
doi = {https://doi.org/10.1016/j.cpc.2020.107275}
}

\end{document}